\documentclass[final,sort&compress]{elsarticle}
\usepackage[english]{babel}
\usepackage[utf8]{inputenc}
\usepackage{xcolor}
\usepackage[plainpages=false,pdfpagelabels,unicode]{hyperref}
\usepackage[margin=2cm]{geometry}

\journal{Applied Surface Science}
\newcommand{\NV}[1]{\mathrm{N}^{{#1}\mathrm{V}_\mathrm{Ti}}}
\newcommand{\TiON}[0]{Ti$_{1-\delta}$O$_x$N$_{1-x}$}

\newcommand{\MCh}[1]{Materials Chemistry, RWTH Aachen University, Kopernikusstr. 10, 52074 Aachen, Germany}
\newcommand{\Uppsala}[1]{Department of Physics and Astronomy, Uppsala University, Lägerhyddsvägen 1, S-75120 Uppsala, Sweden}
\newcommand{\Leoben}[1]{Department of Materials Science, Montanuniversität Leoben, Franz-Josef-Strasse 18, A-8700 Leoben, Austria}

\begin{document}

\author[MCh]{Pavel Ondračka\corref{cor1}}
\ead{pavel.ondracka@gmail.com}
\author[MCh]{Marcus Hans}
\author[MCh]{Damian M. Holzapfel}
\author[Uppsala]{Daniel Primetzhofer}
\author[Leoben]{David Holec}
\author[MCh]{Jochen M. Schneider}
\address[MCh]{\MCh{}}
\address[Uppsala]{\Uppsala{}}
\address[Leoben]{\Leoben{}}
\cortext[cor1]{Corresponding author}

\title{{\it Ab initio}-guided X-ray photoelectron spectroscopy quantification of Ti vacancies in \TiON{} thin films}

\begin{frontmatter}


\ifdefined\rtf{}
\maketitle
\fi

\begin{abstract}
{\it Ab initio} calculations were employed to investigate the effect of oxygen concentration dependent
Ti vacancies formation on the core electron binding energy shifts in cubic titanium oxynitride (\TiON{}).
It was shown, that the presence of a Ti vacancy reduces the 1s core electron binding energy of the first N neighbors by $\sim$0.6\,eV and that
this effect is additive with respect to the number of vacancies.
 Hence it is predicted that the Ti vacancy concentration
can be revealed from the intensity of the shifted components in the N\,1s core spectra region.
This notion was critically appraised by fitting the N\,1s region obtained via
X-ray photoelectron spectroscopy (XPS) measurements of \TiON{} thin films deposited by high power pulsed magnetron sputtering.
A model to quantify the Ti vacancy concentration based on the intensity ratio between the
N 1s signal components,
corresponding to N atoms with locally different Ti vacancy concentration, was developed.
Herein a random vacancy distribution was assumed and the influence of surface oxidation
from atmospheric exposure after deposition was considered.
The so estimated vacancy concentrations are consistent with a model
calculating the vacancy concentration based on the O concentrations determined
by elastic recoil detection analysis and text book oxidation states and hence electroneutrality.
Thus, we have unequivocally established that the formation and population of
Ti vacancies in cubic \TiON{} thin films can be quantified by XPS measurements
from N\,1s core electron binding energy shifts.
\end{abstract}

\end{frontmatter}

\section{Introduction}

It is generally accepted that point defects affect stability and mechanical properties of nitrides and oxynitrides~\cite{Hans2014,Ruess2020,Holzapfel2021}.
Nitrogen vacancies were used previously to increase hardness in TiN$_x$ films~\cite{Lee2005,Shin2003} and toughness in V$_{0.5}$Mo$_{0.5}$N$_x$~\cite{Kindlund2014}.
Another example of point defect engineering are efforts to improve temperature stability in TiAlN~\cite{ToBaben2016}.
Furthermore, it was shown that nitrogen vacancies can be used to stabilize the mechanically unstable cubic MoN~\cite{Koutna2016}, TaN~\cite{Koutna2016} and WN~\cite{Balasubramanian2016}.
Theoretically, point defects are readily described, 
however, precise point defect quantification is a challenging experimental task and is usually
done via indirect estimates such as composition and/or lattice parameter measurements~\cite{ToBaben2016, Hans2014}.
While positron annihilation spectroscopy can quantify defects like vacancies or dislocations,
a knowledge of lifetime for positrons annihilated in the bulk (defect free) material is required~\cite{Gubicza2017}.

However, since it was established that the coordination number of an atom influences the core electron binding energies (BE)~\cite{Bagus1999},
it is possible to detect and quantify the point defect concentration by using X-ray photoelectron spectroscopy (XPS),
provided that the population of point defects and the corresponding energy shift can be resolved spectroscopically.
One material class with inherently high point defect concentrations are rocksalt structure oxynitrides.
It was shown that the presence of oxygen induces metal vacancies in significant concentrations in VAlON~\cite{Shaha2013} and TiAlON~\cite{Schalk2016O, Schalk2016, Hans2014} and more specifically,
that the vacancy concentration depends directly on the oxygen content based on a simple textbook oxidation state electroneutrality model (1 Ti vacancy per 3 O atoms)~\cite{Shaha2013, Hans2014}.
Therefore, the oxynitrides are selected for a critical appraisal of the XPS-based vacancy detection and quantification notion.

In recent years, the importance of {\it ab initio} calculations in analysis and understanding of the
X-ray photoelectron spectroscopy measurements has been growing.
Some recent examples include 
studies of poly-epoxy polymer surface bonding~\cite{Duguet2015, Duguet2019, Gavrielides2016}
 determination of N-doping positions in TiO$_2$~\cite{Panepinto2020},
determination of the structure of epitaxially grown silicene~\cite{Lee2017},
 efforts to distinguish the amount of O in TiO$_2$, mixed TiSi and SiO$_2$-like environment,
allowing quantification of Ti and Si mixing in the Ti$_x$Si$_{1-x}$O$_2$ thin films~\cite{Ondracka2019}
and others~\cite{Aarva2019, Lee2018, Yamazaki2018}.

In this work, we combine {\it ab initio} density functional theory (DFT) calculations and XPS measurements
 to critically appraise if the presence and population of Ti vacancies can be resolved by XPS in  
 \TiON{} thin films with varying O content.

\section{Methodology}

\TiON{} thin films were deposited using reactive 
high power pulsed magnetron sputtering (HPPMS) utilizing a combinatorial setup~\cite{toBabenthesis}.
Si (100) substrates were positioned at a distance of 10\,cm with respect to the cathode.
Oxygen was introduced at the top of the 0.5\,m long cathode to achieve a gradient in the partial oxygen pressure over the cathode length.
Consequently, a chemical composition spread was realized on the substrates as a function of the distance
from the oxygen inlet.
The base pressure was always $<$\,9$\times$10$^{-5}$\,Pa and increased to $<$\,2.5$\times$10$^{-4}$\,Pa after heating the substrates to $\sim$380\,$^\circ\mathrm{C}$.
The Ar (purity $>$ 99.999\,\%) flow was set to 200\,square cubic centimeter per minute (sccm) and N$_2$ (purity $>$ 99.999\,\%) flow was 30\,sccm,
leading to total pressure of approximately 0.43\,Pa for all depositions.
The Ti target (purity $>$ 99\,wt.\,\%) was sputtered at a duty cycle of 2.5\,\% and at a frequency of 500\,Hz (20\,$\mu$s on time, 1950\,$\mu$s off time),
with a time-average power of 1485\,W.
The depositions were done in four batches, using 0, 0.45, 0.9 and 1.35\,sccm of O$_2$ flow (purity $>$ 99.999\,\%) respectively,
leading to peak currents of 229, 229, 219 and 208\,A and peak power densities of 361, 367, 348 and 333 Wcm$^{-2}$ for the respective batches.
The Si substrates were kept at a floating potential.
After 90\,minute depositions, the samples were cooled in vacuum for at least 4 hours and only vented when the sample holder temperature was below 60\,$^\circ$C
in order to minimize the surface oxidation after deposition~\cite{Greczynski2016venting}.

XPS measurements were carried out in a KRATOS AXIS SUPRA spectrometer, using an Al K$_\alpha$ monochromatic source at 1486.6\,eV.
High resolution spectra of the N\,1s region were taken using the pass energy of 10\,eV, energy step of 0.04\,eV, integration time of 1\,s and 12 sweeps.
All of the samples were sufficiently conductive and no charging was observed.
The BE scale of the spectrometer was calibrated using the sputter-cleaned polycrystalline Ag foil,
leading to Ag 3d$_{5/2}$ peak position at 368.2\,eV.
Chemical composition depth profiling was done by time-of-flight elastic recoil detection analysis (ERDA) at the Tandem Laboratory of Uppsala University
and details can be found in~\cite{Azina2020}.
Nitrogen concentrations were corrected based on a stoichiometric TiN reference thin film~\cite{Sortica2017}.
Average chemical compositions were obtained from the depth profiles by excluding the surface-near region.
The statistical uncertainty in all measured depth profiles was $<$\,0.4\,at.\,\%.
In this  manuscript all samples are referred to by their O concentration as measured by ERDA.
X-ray diffraction (XRD) measurements were done with a Bruker D8 Discovery general area detector diffraction
system (GADDS) using Cu(K$\alpha$) radiation.
More detailed information pertaining to experimental procedures employed can be found in the Supplementary material (SM).

Density functional theory~\cite{Hohenberg1964, Kohn1965} was employed for the calculations of BE shifts.
\TiON{} 2$\times$2$\times$2 supercells based on the NaCl-type TiN conventional cubic cell were generated using the special quasi-random structure (SQS) method~\cite{Zunger1990} for the electroneutral configurations,
specifically with 1 Ti vacancy per 3 O atoms. Six different compositions with 0, 3, 6, 9, 12 and 15 O atoms per cell were considered.
Additionally, configurations with 1, 2 and 4 extra Ti vacancies (no longer maintaining the electroneutrality) were generated containing the same number of oxygen atoms as the electroneutral configurations.
Three SQS structures were constructed for every composition and the quantities reported here are the average from the three different cells.
Vienna {\it ab initio} package was used for the full structural relaxation of all cells~\cite{Kresse1996a, Kresse1996b}. 
A core-hole approach was used for the calculation of the binding energies of the core level~\cite{Bagus1999}.
The binding energy was calculated as an energy difference between final state,
where electron from the specific core level was placed into the valence band, and the initial ground state.
This method, although not yielding accurate absolute BE values, is able to provide precise BE differences.
BEs were calculated for N\,1s states of all atoms in the electroneutral structures using the Wien2k full-potential all electron code~\cite{Blaha2020}.
Specific implementation and numerical details of the calculations are described in the SM Section S1
and all calculated data are available under the Creative Commons license in the NOMAD archive~\cite{DFTdata}.

\section{Results and discussion}

All as deposited \TiON{} films with varying O concentrations ranging from 0.4\,at.\,\% to 28.6\,at.\,\% (SM Table~S6),
as determined by ERDA measurements,
exhibit a single cubic phase based on XRD (SM Figure~S3).
The film morphology changes from dense columnar grains for samples with low O content
to a fine-grained nanocrystalline microstructure as more O is incorporated (SM Figure~S7).
Further details and discussion about the film composition and structure can be found in SM Sections S3, S4 and S5.

\begin{figure}
    \includegraphics[width=\linewidth]{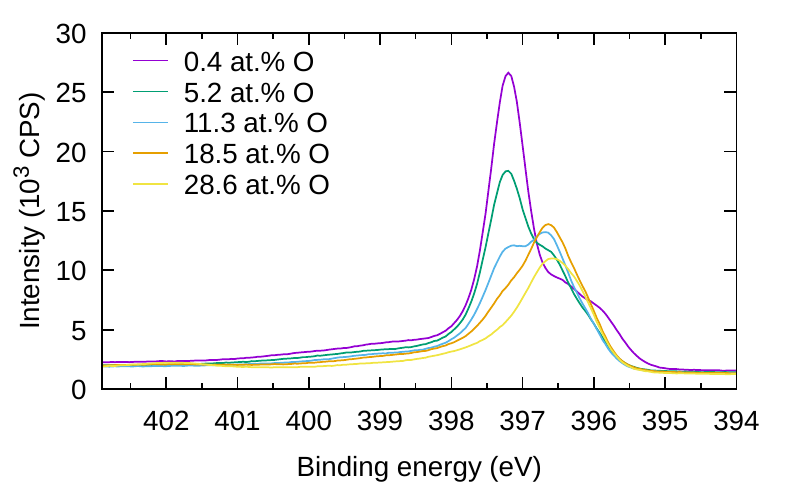}
    \caption{\label{fig:expspectra}Selected measured N\,1s XPS spectra.}
\end{figure}

N\,1s core level XPS measurements of selected samples are presented in Figure~\ref{fig:expspectra}.
As the O content increases, significant shifts of the spectral weights to lower BEs are clearly visible.
Qualitatively, this is consistent with previous reports on TiN surface oxidation,
where the surface oxynitride component was reported at lower binding energies with respect to the main TiN component~\cite{Greczynski2016,Greczynski2016venting}.
The decreasing intensity of the N\,1s signal is caused by the overall decrease of the N content as a consequence of the substitution by O.

\begin{figure}
    \includegraphics[width=\linewidth]{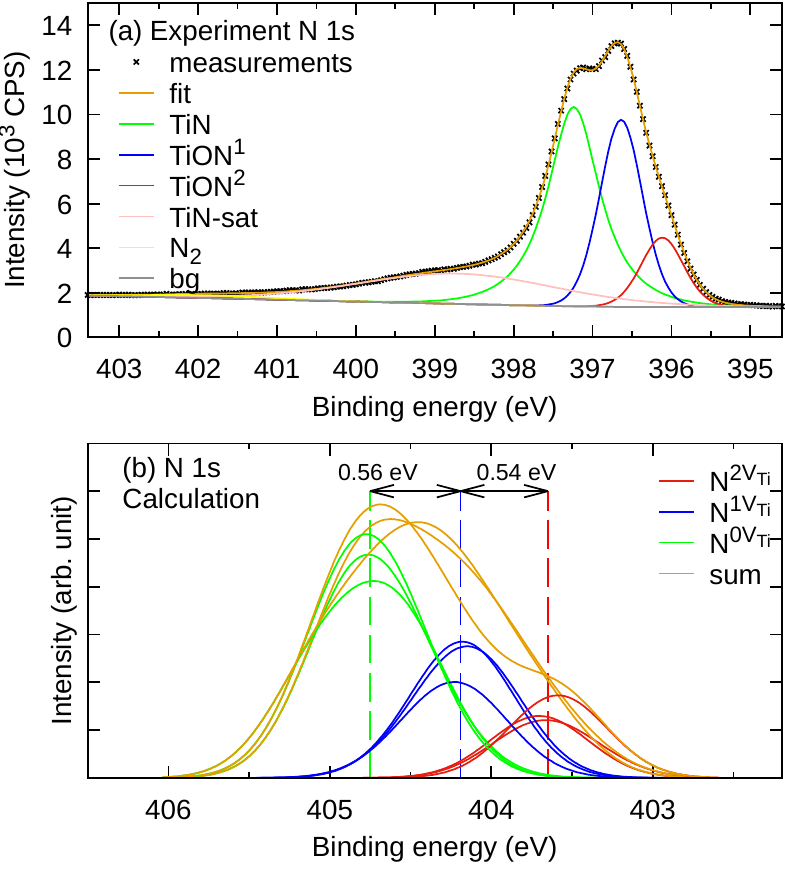}
    \caption{\label{fitXPS}(a) Example fit of the N\,1s spectra for the sample with 11.3\,at.\,\%\,O. Fits of other compositions can be found in SM Figure S9. (b) Calculated N\,1s broadened binding energies for electroneutral \TiON{} with 14.7\,at.\,\% of O. Three lines correspond to three different SQS cells with the same composition. The lines were produced by replacing the calculated discrete binding energies by Gaussians at the same energy position with $\sigma = 0.3$\,eV to produce spectral-like curves for the sole purpose of visual comparison with the experimental spectra. The dashed lines correspond to mean BE values for the specific components. All calculated compositions are shown in SM Figure S2.}
\end{figure}

For fitting of the N\,1s XPS data we modified the approach from Greczynski and Hultman~\cite{Greczynski2016}.
Here, minor differences include a Tougaard-like background~\cite{Tougaard1988}
and the separation of single TiO$_x$N$_y$ component into two components TiON$^1$ and TiON$^2$ was implemented,
as reasonable fits using four peaks only from~\cite{Greczynski2016}
 in the case of TiN-oxidation, namely TiN, TiN-sat, TiO$_x$N$_y$ and N$_2$ could not be obtained.
However, a fitting procedure using an additional TiON peak (TiN, TiON$^1$, TiON$^2$, TiN-sat and N$_2$),
resulted in high quality fits for all studied compositions.
An example fit for one selected composition is presented in Figure~\ref{fitXPS}(a) and shows that all of the components are well distinguishable.
Gaussian-Lorentzian symmetric peak shapes in a product form with 30\,\% Lorentzian ratio were used for four of the fitted components,
with the exception of the TiN component, where again consistently with Greczynski and Hultman~\cite{Greczynski2016} a 90\,\% Lorentzian ratio was used.
Intensity, width and position (mean BE) of the five components were free parameters during the fitting,
while the width of the TiON$^1$ was fixed to be the same as of the TiON$^2$, to reduce the number of free parameters.
It was not possible to fit the position of the N$_2$ component as it was too weak in some compositions leading to ambiguity,
however this was solved by performing a multi-sample fit,
i.e., fitting all the compositions at the same time with the position of the N$_2$ component as a free parameter but shared by all compositions.
This resulted in a value of 401.86\,eV for N$_2$, very close to the reported value of 401.80\,eV from Greczynski and Hultman~\cite{Greczynski2016}.

\begin{figure}
    \includegraphics[width=\linewidth]{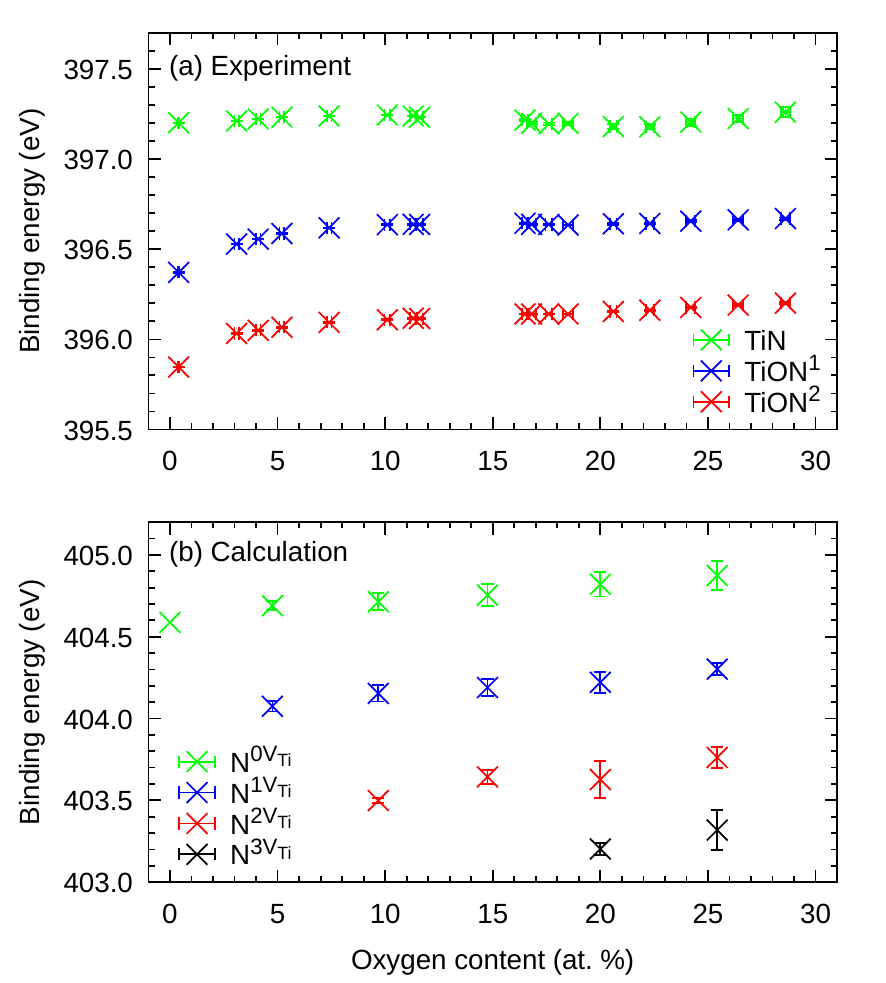}
    \caption{\label{position:evol} Evolution of (a) fitted positions of the TiN, TiON$^1$ and TiON$^2$ components, (b) calculated mean binding energies of $\NV{0}$, $\NV{1}$, $\NV{2}$ and $\NV{3}$ components, as a function of O content.}
\end{figure}

The fitted mean BEs of the TiN, TiON$^1$, and TiON$^2$ components are shown in Figure~\ref{position:evol}(a) and Table~\ref{tab:XPS}.
The BE differences between the TiN and TiON$^1$ BEs are approximately 0.6\,eV for most compositions,
while the differences between the TiON$^1$ and TiON$^2$ BEs are slightly smaller, approximately 0.5\,eV.

\begin{figure}
    \includegraphics[width=0.8\linewidth]{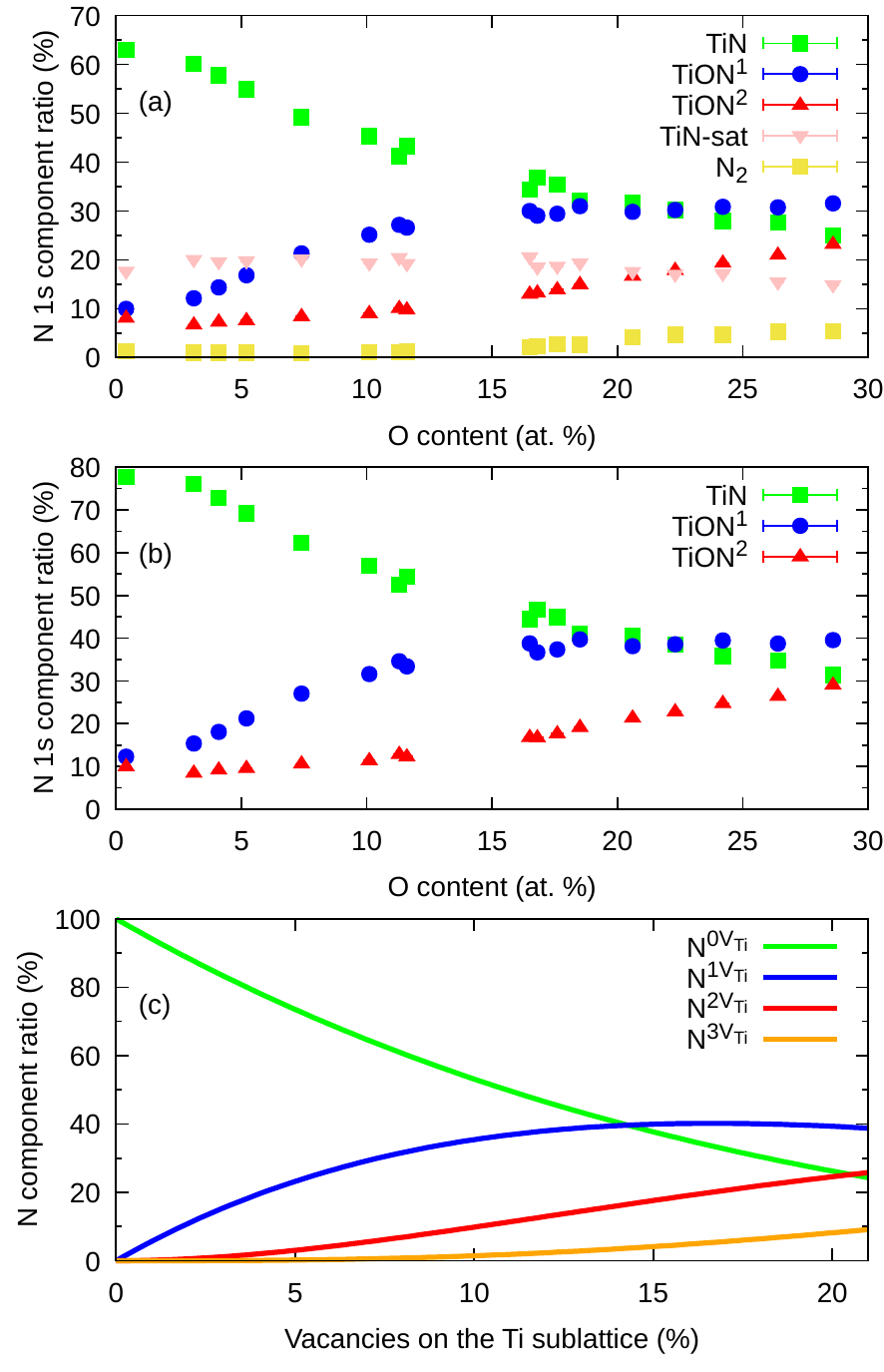}
    \caption{\label{ratiosXPS} a) Oxygen dependent N\,1s component evolution, b) TiN, TiON$^1$, and TiON$^2$ components renormalized to 1, c) theoretical dependency of the $\NV{0}$, $\NV{1}$, and $\NV{2}$ components, under the assumption of random vacancy distribution.}
\end{figure}

\begin{table*}
\begin{center}
\caption{\label{tab:XPS}Fitted XPS N\,1s component ratios and mean BEs for all samples as well as the mean BE differences between the TiN, TiON$^1$ and TiON$^2$ components.}
\begin{tabular}{lcccccccccccc}
\hline\hline
[O] & $f^\mathrm{TiN}$ & $f^\mathrm{TiON^1}$ & $f^\mathrm{TiON^2}$ & $f^\mathrm{sat}$ & $f^\mathrm{N_2}$ & $E^\mathrm{TiN}$ & $E^\mathrm{TiON^1}$ & $E^\mathrm{TiON^2}$ & $E^\mathrm{sat}$ & $E^\mathrm{N_2}$ & $E^\mathrm{TiN}$ & $E^\mathrm{TiON^2}$  \\
 & & & & & & & & & & & $- E^\mathrm{TiON^1}$ & $-E^\mathrm{TiON^1}$ \\
(at.\,\%) & (\%) & (\%) & (\%) & (\%) & (\%) & (eV) & (eV) & (eV) & (eV) & (eV) & (eV) & (eV) \\
\hline
0.4 & 63.0 & 10.0 & 8.1 & 17.6 & 1.3 & 397.20 & 396.37 & 395.85 & 398.97 & 401.86 & 0.83 & 0.52 \\
3.1 & 60.2 & 12.1 & 6.7 & 20.0 & 1.0 & 397.21 & 396.53 & 396.03 & 398.92 & 401.86 & 0.68 & 0.50 \\
4.1 & 57.8 & 14.3 & 7.3 & 19.6 & 1.0 & 397.22 & 396.56 & 396.05 & 398.90 & 401.86 & 0.67 & 0.51 \\
5.2 & 54.9 & 16.8 & 7.6 & 19.7 & 0.9 & 397.23 & 396.59 & 396.07 & 398.88 & 401.86 & 0.64 & 0.52 \\
7.4 & 49.2 & 21.3 & 8.4 & 20.2 & 0.9 & 397.24 & 396.62 & 396.09 & 398.82 & 401.86 & 0.62 & 0.53 \\
10.1 & 45.3 & 25.2 & 9.1 & 19.3 & 1.1 & 397.24 & 396.64 & 396.11 & 398.81 & 401.86 & 0.61 & 0.53 \\
11.3 & 41.2 & 27.2 & 10.1 & 20.4 & 1.1 & 397.24 & 396.64 & 396.12 & 398.71 & 401.86 & 0.60 & 0.52 \\
11.6 & 43.3 & 26.6 & 9.8 & 19.2 & 1.2 & 397.23 & 396.64 & 396.12 & 398.77 & 401.86 & 0.60 & 0.52 \\
16.5 & 34.4 & 30.0 & 13.0 & 20.6 & 2.0 & 397.22 & 396.64 & 396.14 & 398.53 & 401.86 & 0.57 & 0.50 \\
16.8 & 36.9 & 29.0 & 13.3 & 18.5 & 2.3 & 397.20 & 396.64 & 396.14 & 398.63 & 401.86 & 0.56 & 0.50 \\
17.6 & 35.4 & 29.4 & 13.9 & 18.6 & 2.7 & 397.19 & 396.64 & 396.14 & 398.56 & 401.86 & 0.56 & 0.50 \\
18.5 & 32.1 & 31.0 & 15.0 & 19.3 & 2.6 & 397.20 & 396.63 & 396.14 & 398.41 & 401.86 & 0.56 & 0.49 \\
20.6 & 31.7 & 29.8 & 16.7 & 17.6 & 4.2 & 397.18 & 396.64 & 396.15 & 398.40 & 401.86 & 0.54 & 0.49 \\
22.3 & 30.2 & 30.2 & 17.9 & 17.0 & 4.7 & 397.18 & 396.64 & 396.16 & 398.37 & 401.86 & 0.54 & 0.48 \\
24.2 & 28.0 & 30.9 & 19.3 & 17.2 & 4.7 & 397.20 & 396.66 & 396.18 & 398.30 & 401.86 & 0.55 & 0.48 \\
26.4 & 27.6 & 30.7 & 21.0 & 15.4 & 5.2 & 397.22 & 396.66 & 396.19 & 398.36 & 401.86 & 0.56 & 0.47 \\
28.6 & 25.0 & 31.6 & 23.2 & 14.8 & 5.4 & 397.26 & 396.67 & 396.20 & 398.33 & 401.86 & 0.59 & 0.47 

\\
\hline
\multicolumn{11}{r}{mean difference:} & 0.60 & 0.50 \\
\hline\hline
\end{tabular}
\end{center}
\end{table*}

\begin{figure*}
    \includegraphics[width=\linewidth]{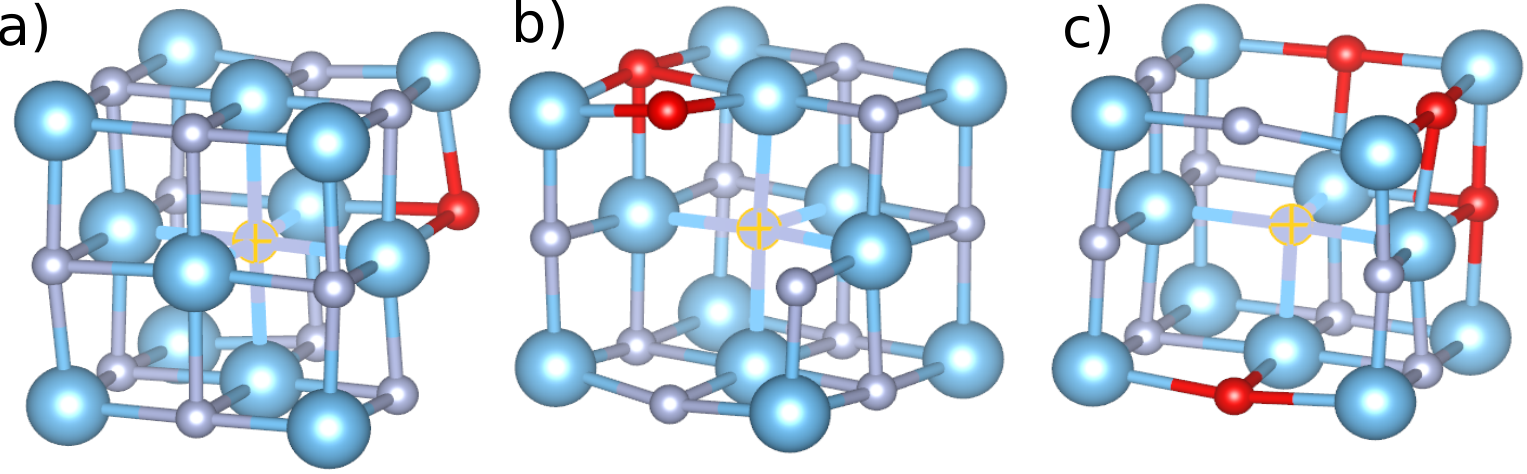}
    \caption{\label{fig:vac-scheme} Example of a) $\NV{0}$, b) $\NV{1}$ and c) $\NV{2}$ atoms (crossed atom at the center). N atoms are gray, Ti are blue and O atoms are marked red. Visualized by VESTA~\cite{Momma2011}.}
\end{figure*}

DFT calculations were employed to critically appraise the working hypothesis of this paper,
namely that the presence of Ti vacancies causes significant and hence measurable
changes in the N\,1s core level binding energies and that subsequently the population of defects can be quantified by XPS.
For evaluation of the N\,1s core levels, the calculated binding energies of all N atoms in the structure were hence divided into groups based on the number
of Ti vacancies in the first coordination shell.
Nitrogen with 6 Ti neighbors and 0 Ti vacancies is being labeled as $\NV{0}$,
while $\NV{1}$ has 5 Ti neighbors and 1 Ti vacancy, etc., see Figure~\ref{fig:vac-scheme}.
Figure~\ref{fitXPS}(b) shows a broadened calculated binding energies of the N\,1s core levels for the electroneutral \TiON{} with 14.8\,at.\,\% of O. 
The binding energies for $\NV{1}$ and $\NV{2}$ atoms are shifted to smaller binding energies by 0.56\,eV and 1.1\,eV with respect to
the main $\NV{0}$ component, respectively.
The evolution of mean $\NV{0}$, $\NV{1}$, $\NV{2}$ and $\NV{3}$ N\,1s binding energies as a function of O content for all calculated compositions
is shown in Figure~\ref{position:evol}(b).
The peak positions of the individual $\NV{n}$ components shift to a slightly higher BEs as the oxygen content increases,
however the overall shift is below 0.2\,eV over the whole oxygen concentration range.
The shifts between the different $\NV{n}$ components corresponding to
the different numbers of vacancies are, however, very similar for all compositions.
This suggests that the relative BE shifts do not depend on the overall composition of the films,
but are sensitive to the specific local chemical environment.
Furthermore, it is evident that the effect of the Ti vacancies on the N\,1s core level is additive and amounts to
0.5--0.6\,eV BE decrease per Ti vacancy in the first coordination shell.
Therefore, the calculated binding energies predict a significant shift of the N\,1s core level,
due to the presence of Ti vacancies in the first coordination shell
which is consistent with our working hypothesis that the presence and population
of Ti vacancies in \TiON{} thin films can be quantified by the XPS measurements.

The relative intensity ratios of the five fitted components for all films are shown in the Figure~\ref{ratiosXPS}(a) and Table~\ref{tab:XPS}.
The relative intensity ratio of the TiN component is the dominating one up to $~\sim$22\,at.\,\% O content and is decreasing with increasing O content over the whole composition range.
The relative intensity ratio of the TiON$^1$ component increases up to $\sim$30\,\% at 18.5\,at.\,\% O and saturates at this level for higher oxygen concentrations.
Similarly, the relative intensity ratio of the TiON$^2$ component is also increasing with the O content over the investigated composition range.
However, contrary to the TiON$^1$ the increase of the TiON$^2$ component is moderate at low O contents and then increases monotonically above 15\,at.\,\% O.
The intensity ratio of the satellite peak is $\sim$20\,\%, similar for all here probed samples. 
The N$_2$ component is negligible in films with low O content and only becomes apparent in films with more than 15\,at.\,\% O,
however its relative intensity is only around 5\,\% even in the O-rich films.
The intensities of the TiN, TiON$^1$ and TiON$^2$ components are renormalized to sum to 1 and shown in Figure~\ref{ratiosXPS}(b).

The obtained relative intensity versus O concentration trends are in a very good agreement with the theoretical expectations depicted in Figure~\ref{ratiosXPS}(c) for the relative intensities of
the $\NV{n}$ components as a function of the amount of vacancies on the metal sublattice $\delta$, where, assuming a random vacancy distribution,
the probability of the N atom to have $n$ Ti vacancies in the first coordination shell is
\begin{equation}
P^{\NV{n}}(\delta) = {6 \choose {n}} \delta^{n} (1-\delta)^{6-n} \mathrm{.}
\label{prob}
\end{equation}
The good agreement between the theoretical prediction and observed trends together with
the very good agreement between the experimental and theoretical BE shifts is a direct evidence
that the TiN, TiON$^1$ and TiON$^2$
components correspond to N atoms with 0, 1 and 2 close Ti vacancies respectively.

A discrepancy between prediction and experiment is visible at small O contents.
If TiN, TiON$^1$ and TiON$^2$ components
indeed directly correspond to $\NV{0}$, $\NV{1}$ and $\NV{2}$ components,
their relative intensity ratios should be 1, 0 and 0 for pure TiN respectively,
which is not the case.
This disagreement is caused by the surface oxidation from atmospheric exposure. Greczynski and Hultman~\cite{Greczynski2016}
report the surface oxynitride peak to be shifted by 1.05\,eV to lower BEs from the main TiN peak,
which causes an overestimation of the $\NV{2}$ and potentially of the $\NV{1}$ component.
This is visible also in Figure~\ref{fitXPS}(a) and (b).
At similar O contents the experimental TiON$^1$ and TiON$^2$ components are much stronger than the 
$\NV{1}$ and $\NV{2}$ calculated components, due to the additional surface oxidation in the experiment.
In fact, it was previously reported that Ti vacancies are formed at the surface during oxidation of TiN~\cite{Zimmermann2009}
which is consistent with the findings presented herein.

There is no direct evidence for the presence of a $\NV{3}$ component in the experimental data which
is likely caused by its relatively small population:
assuming the random vacancy distribution, the $\NV{3}$ ratio is less than 10\,\% at 20\,\% metal vacancies on the metal sublattice.
Furthermore, an energy penalty of 0.68\,eV and 0.31\,eV was reported for close Ti vacancies on (110) and (100) chains~\cite{Tsetseris2007}.
This energetic penalty further reduces the ratio of the $\NV{3}$ component at low Ti vacancy concentrations.

Before any quantification is possible, the influence of the surface oxidation from the atmospheric exposure has to be corrected for.
Greczynski {\it et al.}~\cite{Greczynski2016venting} showed, that for pure TiN, four regions at the surface are present: the carbon contamination on top, pure oxide layer and a transition oxynitride layer which evolve due to atmosphere exposure and a bulk-like nitride (oxynitride in our case) region at the bottom.
Since the N\,1s peak is used to estimate the vacancy concentrations, it is possible to ignore the surface oxide layer as there is no nitrogen present,
however the influence of the oxynitride transition layer needs to be taken into account,
otherwise the amount of vacancies will be grossly overestimated.

The here applied correction procedure rests on the assumption that the oxidation
of all samples is similar as the venting procedure and sample transport to the XPS were similar for all samples.
Hence, it is reasonable to assume that the ratio of the nitrogen in the transitional oxynitride region to the nitrogen in the bulk-like region is constant for all of the samples.
Therefore, we fit the component curves as shown in the Figure~\ref{ratiosXPS}(b) with functions similar to Eq. (\ref{prob}).
However, a constant offset was applied to account for signal contributions due to surface oxidation by atmospheric exposure.

Since the experimental $\NV{0}$, $\NV{1}$ and $\NV{2}$ are normalized, the fitting functions are normalized as well
\begin{equation}
f^{\NV{0}}_\mathrm{norm}(\delta) = \frac{P^{\NV{0}}(\delta)}{P^{\NV{0}}(\delta) + P^{\NV{1}}(\delta) + P^{\NV{2}}(\delta)} - C_1 - C_2 \mathrm{,} \\
\end{equation}
\begin{equation}
f^{\NV{1}}_\mathrm{norm}(\delta) = \frac{P^{\NV{1}}(\delta)}{P^{\NV{0}}(\delta) + P^{\NV{1}}(\delta) + P^{\NV{2}}(\delta)} + C_1 \mathrm{,} \\
\end{equation}
\begin{equation}
f^{\NV{2}}_\mathrm{norm}(\delta) = \frac{P^{\NV{2}}(\delta)}{P^{\NV{0}}(\delta) + P^{\NV{1}}(\delta) + P^{\NV{2}}(\delta)} + C_2 \mathrm{,}
\end{equation}
where $C_1$ and $C_2$ take the effect of surface oxidation due to atmosphere exposure into account.
It is furthermore assumed during the fitting of $C_1$ and $C_2$ the relationship between vacancy concentration and oxygen content can be described by a constant c,
$\delta = c [\mathrm{O}]$, a free parameter in the fit.
The resulting fitted values of $C_1$ and $C_2$ are 5.3\,\% and 8.3\,\%.

With the component ratios corrected for the surface oxidation due to atmospheric exposure,
the vacancy population can be estimated.
Using the three TiN, TiON$^1$ and TiON$^2$ components, it is possible to obtain not only the vacancy concentration,
but also their distribution,
i.e., information about clustering or separation of the vacancies.
However, for simplicity an averaging scheme for the values obtained from the separate components is used, while still assuming the random distribution.
This approach searches for a $\delta$, for which squared weighted differences between the theoretical $\NV{0}$, $\NV{1}$, $\NV{2}$ and experimental TiN, TiON$^1$, TiON$^2$ ratios is minimized,
i.e., to find the minimum of a following function
\begin{equation}
g(\delta) = \left(f^{\NV{0}}_\mathrm{norm}(\delta) - f^\mathrm{TiN}\right)^2\left(\frac{1}{\sigma^\mathrm{TiN}}\right)^2 
+ \left(f^{\NV{1}}_\mathrm{norm}(\delta) - f^{\mathrm{TiON}^1}\right)^2\left(\frac{1}{\sigma^{\mathrm{TiON}^1}}\right)^2
+ \left(f^{\NV{2}}_\mathrm{norm}(\delta) - f^{\mathrm{TiON}^2}\right)^2\left(\frac{1}{\sigma^{\mathrm{TiON}^2}}\right)^2
\, ,
\end{equation}
where $f^\mathrm{TiN}$, $f^{\mathrm{TiON}^1}$, $f^{\mathrm{TiON}^2}$ and $\sigma^\mathrm{TiN}$, $\sigma^{\mathrm{TiON}^1}$, $\sigma^{\mathrm{TiON}^2}$ are the fitted component ratios and their uncertainties for a specific O concentration.

\begin{figure}
    \includegraphics[width=\linewidth]{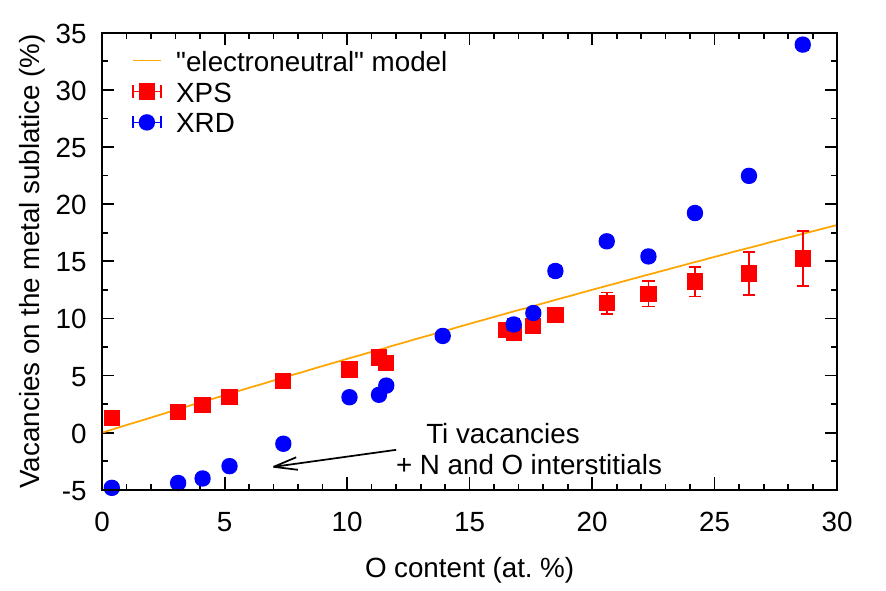}
    \caption{\label{vac}Vacancy concentrations as estimated using the here developed XPS method. Simple estimate based on XRD lattice parameters is added for comparison, as well as a theoretical line for perfectly electroneutral composition (assuming +3, -3 and -2 oxidation number for Ti, N and O respectively). All of the models assume that the vacancies are the only defects present in the films. This is not perfectly satisfied below 13\,at.\,\% O. See SM Section S4 and S7 for more details.}
\end{figure}

The vacancy concentration estimates using the previously described method are summarized in Figure~\ref{vac}.
The results show a linear increase over the whole oxygen composition range,
from 1.3\,\% of vacancies on the metal sublattice at 0.4\,at.\,\% of O to 15.2\,\% of vacancies on the metal sublattice 
at 28.6\,\%. of O and are in a perfect agreement with the simple electroneutral model.
The simpler estimate of the composition dependent vacancy concentrations
based on XRD lattice parameter replicate the trend but exhibits local
deviations to both the electroneutral model
as well as the composition dependent vacancy model based on the XPS data.
While potential causes for the aforementioned local deviations are discussed in SM Section~7,
the here presented results confirm that the Ti vacancies are inherently present in and stabilize the cubic titanium oxynitride and 
give further support to the previously proposed simple electroneutrality model~\cite{Shaha2013, Hans2014}.
Furthermore it is evident that the population of Ti vacancies in \TiON{} for up to 28.6\,at.\,\% O
can be quantified with the here proposed model utilizing binding energy data obtained from the N\,1s core spectra region via XPS.

\section{Conclusions}

Based on the {\it ab initio} and spectroscopy data, 
we have unequivocally established that the N\,1s XPS signal of \TiON{} is sensitive to the presence of Ti vacancies
and that the corresponding vacancy concentration can be quantified with the here proposed model.
Spectroscopic analysis of the N\,1s region of \TiON{} thin films with systematically varied oxygen concentration
and therefore systematically varied Ti vacancy concentration revealed two components at approximately
0.6\,eV and 1.1\,eV lower binding energies with respect to the main TiN-like component.
Using {\it ab initio} DFT calculations we established that these binding energy shifts are caused
by the presence of 1 and 2 Ti vacancies in the first coordination shell of N atoms respectively.
A theoretical model was proposed to quantify the vacancy concentration based on the relative ratios of the three N\,1s components.
As surface oxidation of the \TiON{} films also affect the N\,1s binding energy,
the corresponding signal contribution from the oxidized surface was taken into account.
The so estimated XPS vacancy concentrations are consistent with
the theoretical electroneutral model based on textbook oxidation states.

\section*{CRediT authorship contribution statement}
Pavel Ondračka: Conceptualization, Formal analysis, Investigation, Methodology,
Writing - original draft, review \& editing, Visualization.
Marcus Hans: Formal analysis, Investigation, Writing - review \& editing, Visualization.
Damian M. Holzapfel: Formal analysis, Investigation, Writing - review \& editing, Visualization.
Daniel Primetzhofer: Formal analysis, Investigation, Writing - review \& editing, Funding acquisition.
David Holec: Conceptualization, Supervision, Writing - review \& editing.
Jochen M. Schneider: Conceptualization, Supervision, Project administration, Writing - review \& editing, Funding acquisition.

\section*{Acknowledgments}
This research was funded by German Research Foundation (DFG, SFB-TR 87/3) ''Pulsed high power plasmas for the synthesis of nanostructured functional layer''.
The authors acknowledge financial support from the Swedish research council, VR-RFI (contracts \#821-2012-5144 \& \#2017-00646\_9), and the Swedish Foundation for Strategic Research (SSF, contract RIF14-0053) for the operation of the tandem accelerator at Uppsala University.
The authors also gratefully acknowledge the computing time granted by the JARA Vergabegremium and provided
 on the JARA Partition part of the supercomputer CLAIX at RWTH Aachen University (project JARA0151).

\bibliography{bib-db}

\end{document}


\author[a]{Pavel Ondračka}
\author[a]{Marcus Hans}
\author[a]{Damian M. Holzapfel}
\author[b]{Daniel Primetzhofer}
\author[c]{David Holec}
\author[a]{Jochen M. Schneider}
\affil[a]{Materials Chemistry, RWTH Aachen University, Kopernikusstr. 10, 52074 Aachen, Germany}
\affil[b]{Department of Physics and Astronomy, Uppsala University, Lägerhyddsvägen 1, S-75120 Uppsala, Sweden}
\affil[c]{Department of Materials Science, Montanuniversität Leoben, Franz-Josef-Strasse 18, A-8700 Leoben, Austria}

\title{{\it Ab initio}-guided X-ray photoelectron spectroscopy quantification of Ti vacancies in \TiON{} thin films -- Supplementary material}


\maketitle

\tableofcontents

\section{{\it Ab initio} calculations}

\begin{figure}
\begin{center}
\includegraphics[width=0.8\linewidth]{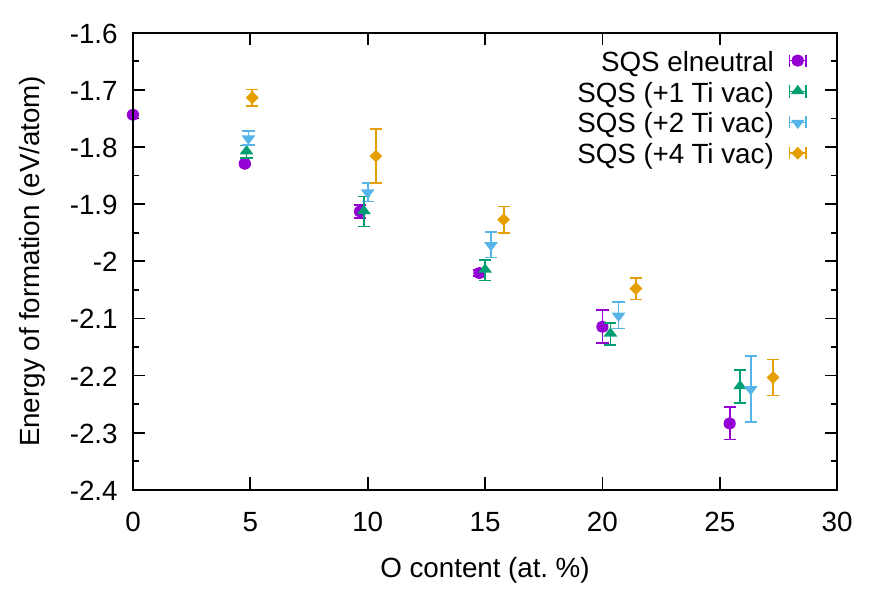}
\caption{\label{fig:Ef}Calculated mean formation energies for TiON SQS cells. Error bars show the standard deviation of the values obtained for the three different SQS cells with the same composition.}
\end{center}
\end{figure}

2$\times$2$\times$2 \TiON{} supercells based on the c-TiN conventional cell were used for the calculations with 0, 3, 6, 9, 12 and 15 oxygen atoms respectively replacing nitrogen on the non-metal sublattice.
According to the electroneutral model~\cite{Shaha2013, Hans2014}, a one third of this amount of vacancies (0, 1, 2, 3, 4, and 5) was generated on the Ti sublattice.
The SQS method~\cite{Zunger1990} was employed to generate a random-like distribution of the Ti and vacancies on the Ti sublattice and N and O on the N sublattice.
The SQS method does not prescribe how the quasi-randomness should be coupled between the metal and non-metal sublatice.
However, three cells were generated for every composition, with different random relations between the metal and nonmetal sublattices. In the cases where only a single value of some property is shown later (like the formation energy in Figure~\ref{fig:Ef}), it was obtained as an average of results from the different cells and is presented together with the standard deviation of the three values.
Besides the electroneutral compositions, additional structures were generated with identical oxygen and nitrogen contents but with 1, 2 and 4 extra Ti vacancies using the same methodology.

The generated cells were fully relaxed with respect to the atomic positions, cell size and shape using the Vienna {\it Ab initio} Simulation Package (VASP)~\cite{Kresse1996a, Kresse1996b}
and the projector augmented-wave potentials~\cite{Blochl1994,Kresse1999}.
A $5 \times 5 \times 5$ Monkhorst-Pack $k$-point grid~\cite{Monkhorst1976}, 500\,eV cutoff energy,
and a Gaussian smearing with the broadening parameter $\sigma$ of 0.2\,eV were employed.
The stopping criterion for the relaxation was that all residual forces are below 0.01\,eV/\r{A},
while the required convergence for the self-consistent loop was 0.1\,$\mu$eV per supercell.

Wien2k all-electron full-potential linearized augmented planewave code~\cite{Blaha2020} was then used to calculate the core electron binding energies (BEs)
for the Ti\,2p$_{3/2}$, N\,1s and O\,1s core levels for every atom in every electroneutral composition.
Single electrons were removed from the specific core levels and placed into the valence band (all calculated compositions are metallic)
and the binding energy was calculated as the energy difference between this final state and the initial ground state.
Atomic spheres radii of $\sim$1.65, $\sim$1.75 and $\sim$1.9\,Bohr radius were used for N, O and Ti, respectively.
7.5\,$R_\mathrm{mt}K_\mathrm{max}$ basis size parameter, together with $4 \times 4 \times 4$ Monkhorst-Pack $k$-point grid~\cite{Monkhorst1976},
Fermi smearing of 25\,meV, and 0.14\,meV total energy convergence criterion
should guarantee a better than 0.1\,eV absolute BE convergence, with even a better convergence of the relative BE shifts.
For the sole purpose of visualization and comparison with experimental spectra,
the calculated discrete binding energies were replaced with Gaussians with a standard deviation of 0.3\,eV.

Energy of formation, $E_\mathrm{f}$, was calculated as
\begin{equation}
E_\mathrm{f}(\mathrm{TiON}^i) = E_0(\mathrm{TiON}^i) - [\mathrm{O}]^i E_0(\mathrm{O_2}) - [\mathrm{N}]^i E_0(\mathrm{N_2}) - [\mathrm{Ti}]^i E_0(\mathrm{Ti}) \, ,
\end{equation}
where $E_0(\mathrm{TiON}^i)$, $E_0(\mathrm{O_2})$, $E_0(\mathrm{N_2})$, and $E_0(\mathrm{Ti})$
are the ground state energies per atom of the specific $\mathrm{TiON}^i$ cell ($i$ marks here a single specific composition), O$_2$ molecule, N$_2$ molecule and hcp-Ti.
$[\mathrm{O}]^i$, $[\mathrm{N}]^i$, and $[\mathrm{Ti}]^i$ are the concentrations of O, N and Ti atoms for the given composition $\mathrm{TiON}^i$.

The calculated energies of formation are shown in Figure~\ref{fig:Ef} and support the electroneutral model,
i.e., the structures with extra vacancies have in general less negative energies than the electroneutral ones at the given O content. 

\begin{sidewaysfigure}
\includegraphics[width=\linewidth]{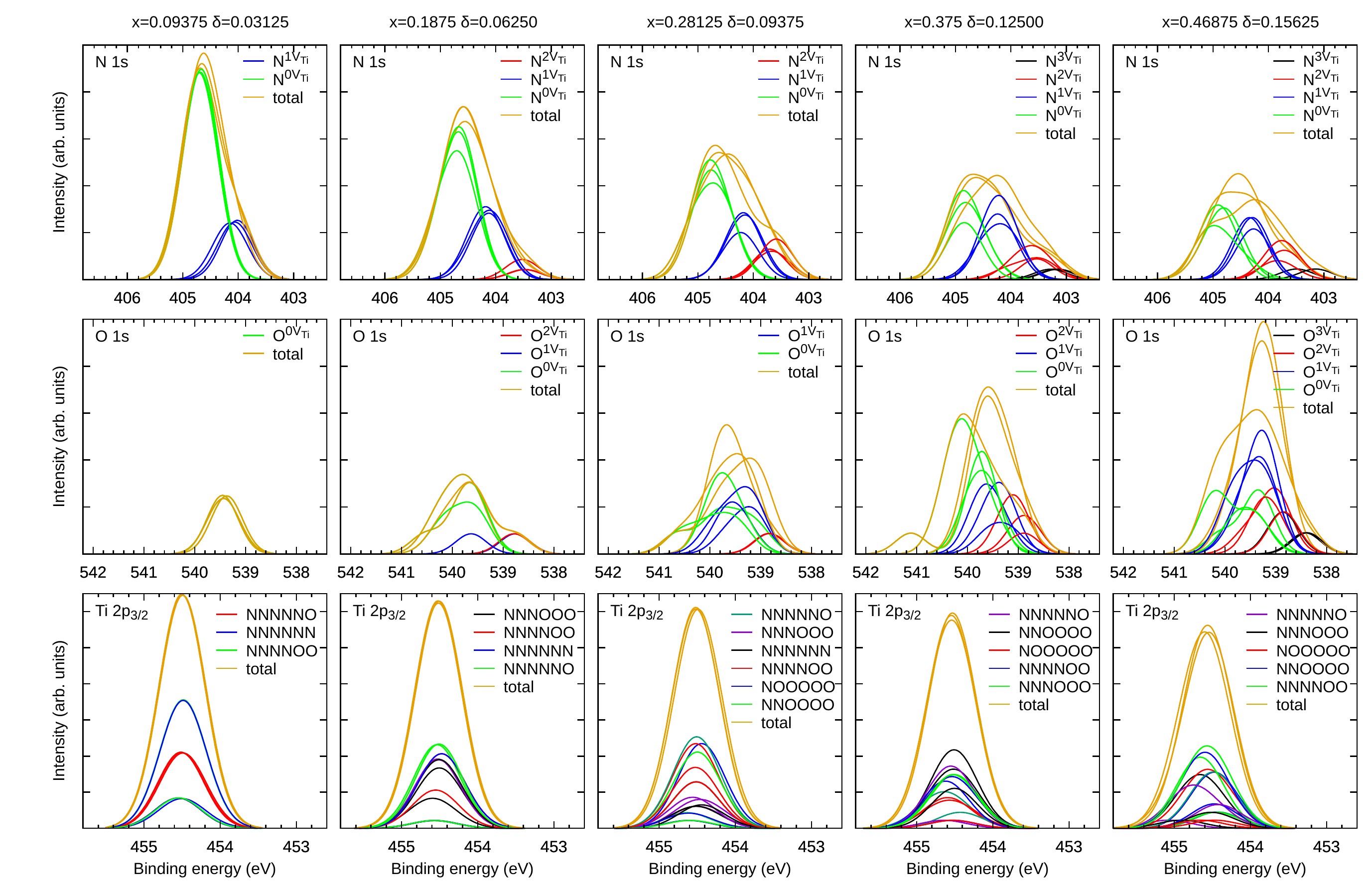}
\caption{\label{fig:spectra}Simulated broadened binding energies spectra of N 1s, O 1s and Ti 2p$_{3/2}$ peak in electroneutral \TiON{}.}
\end{sidewaysfigure}

The calculated broadened binding energies of all electroneutral compositions are shown in Figure~\ref{fig:spectra}.
We note that besides the vacancy induced shifts of BE for N\,1s as discussed in the manuscript,
a similar trend can be found also for the oxygen atoms (Figure~\ref{fig:spectra} middle row),
however the O\,1s binding energies are more scattered, likely due to larger structural disorder in the local O environment
or because of the smaller sample size.
Moreover, the O peak is generally not suitable for Ti vacancy quantification as there is usually a significant oxygen contamination from the environment.
For the Ti atoms which have a constant amount of 6 nearest neighbors, the influence of the specific neighbor configuration on the Ti\,2p$_{3/2}$ binding energy was evaluated as well.
The difference is however negligible as visible in the bottom row in Figure~\ref{fig:spectra}.

\clearpage{}

\section{Depositions}

\begin{table}[h!]
\begin{center}
\caption{\label{dep-table} Detailed deposition conditions}
\begin{tabular}{lcccc}
\hline\hline
internal batch number & 3395 & 3396 & 3397 & 3398 \\
base pressure at RT (10$^{-7}$ mbar)  & 8.4 & 6.6 & 6.4 & 6.4 \\
outer heater power (kW) & \multicolumn{4}{c}{4.5} \\
inner heater power (kW) & \multicolumn{4}{c}{1.5} \\
heating time (h) & \multicolumn{4}{c}{2} \\
temperature after heating ($^\circ$C) & 371 & 380 & 382 & 376 \\
base pressure after heating (10$^{-6}$ mbar) & 2.4 & 2.3 & 2.2 & 2.5 \\
Ar flow (sccm) & \multicolumn{4}{c}{200} \\
pressure with Ar (mPa) & 375 & 373 & 373 & 371 \\
N$_2$ flow (sccm) & \multicolumn{4}{c}{30} \\
pressure with Ar+N$_2$ (mPa) & 428 & 430 & 429 & 427 \\
O$_2$ flow (sccm) & 0 & 0.45 & 0.9 & 1.35 \\
pressure with Ar+N$_2$+O$_2$ (mPa) & 428 & 429 & 429 & 428 \\
target to sample distance (cm) & \multicolumn{4}{c}{10} \\ 
HMMPS set power (W) & \multicolumn{4}{c}{1500} \\
HPPMS on time ($\mu$s) & \multicolumn{4}{c}{50} \\ 
HPPMS off time ($\mu$s) & \multicolumn{4}{c}{1950} \\ 
HPPMS frequency (Hz) & \multicolumn{4}{c}{500} \\ 
HPPMS duty cycle (\%) & \multicolumn{4}{c}{2.5} \\ 
deposition time (min) & \multicolumn{4}{c}{90} \\ 
cooling time (h) & 4.5 & 4 & 5.5 & 5.3 \\
venting temperature ($^\circ$C) & 54 & 60 & 49 & 50 \\
\hline\hline
\end{tabular}
\end{center}
\end{table}

\begin{table}
\begin{center}
\caption{\label{dep-3395} Conditions during deposition run with 0\,sccm O$_2$ flow.}
\begin{tabular}{ccccccc}
\hline\hline
time & pressure& mean power & voltage & mean current & peak current & temperature \\
(min) & (mPa) & (W) & (V) & (A) & (A) & ($^\circ$C) \\
\hline
1 & 425 & 1485 & 698 & 2 & 229 & 376 \\
5 & 428 & 1485 & 697 & 2 & 229 & 379 \\
30 & 431 & 1485 & 694 & 2 & 229 & 385 \\
60 & 432 & 1485 & 692 & 2 & 229 & 388 \\
89 & 433 & 1485 & 691 & 2 & 229 & 389 \\
\hline\hline
\end{tabular}
\end{center}
\end{table}

\begin{table}
\begin{center}
\caption{\label{dep-3396} Conditions during deposition run with 0.45\,sccm O$_2$ flow.}
\begin{tabular}{ccccccc}
\hline\hline
time & pressure& mean power & voltage & mean current & peak current & temperature \\
(min) & (mPa) & (W) & (V) & (A) & (A) & ($^\circ$C) \\
\hline
1 & 426 & 426 & 1485 & 2 & 229 & 381 \\
5 & 427 & 427 & 1485 & 2 & 229 & 382 \\
22 & 430 & 430 & 1485 & 2 & 229 & 388 \\
70 & 431 & 431 & 1486 & 2 & 229 & 390 \\
89 & 432 & 432 & 1485 & 2 & 229 & 391 \\
\hline\hline
\end{tabular}
\end{center}
\end{table}

\begin{table}
\begin{center}
\caption{\label{dep-3397} Conditions during deposition run with 0.9\,sccm O$_2$ flow.}
\begin{tabular}{ccccccc}
\hline\hline
time & pressure& mean power & voltage & mean current & peak current & temperature \\
(min) & (mPa) & (W) & (V) & (A) & (A) & ($^\circ$C) \\
\hline
1 & 426 & 1485 & 705 & 2 & 219 & 384 \\
5 & 428 & 1485 & 702 & 2 & 219 & 387 \\
30 & 429 & 1485 & 701 & 2 & 219 & 388 \\
60 & 429 & 1485 & 701 & 2 & 219 & 389 \\
89 & 431 & 1485 & 701 & 2 & 219 & 389 \\
\hline\hline
\end{tabular}
\end{center}
\end{table}

\begin{table}
\begin{center}
\caption{\label{dep-3398} Conditions during deposition run with 1.35\,sccm O$_2$ flow.}
\begin{tabular}{ccccccc}
\hline\hline
time & pressure& mean power & voltage & mean current & peak current & temperature \\
(min) & (mPa) & (W) & (V) & (A) & (A) & ($^\circ$C) \\
\hline
1 & 426 & 1485 & 709 & 1.9 & 208 & 382 \\
5 & 427 & 1485 & 706 & 2 & 208 & 385 \\
32 & 430 & 1485 & 706 & 2 & 208 & 386 \\
65 & 431 & 1485 & 705 & 2 & 208 & 387 \\
89 & 432 & 1485 & 705 & 2 & 208 & 387 \\
\hline\hline
\end{tabular}
\end{center}
\end{table}

\clearpage{}

\section{ERDA}

The elemental compositions of the films as determined by ERDA for all films is shown in Table~\ref{ERDA-table}. Besides Ti, O and N, traces of C were detected in the films as well,
however the detected carbon content was below 0.5\,at.\,\% in all samples and therefore neglected.
\begin{table}[h!]
\begin{center}
\caption{\label{ERDA-table}ERDA compositions}
\begin{tabular}{ccccccccc}
\hline\hline
O$_2$ flow (sccm) & distance from gas shower (cm) & [Ti] (at.\,\%) & [O] (at.\,\%)  & [N] (at.\,\%) & $x$ \\ 
\hline
0.00 & 30.0 & 50.5 $\pm$ 0.3 & 0.4 $\pm$ 0.0 & 49.1 $\pm$ 0.3 & 0.008 \\
0.45 & 30.0 & 50.1 $\pm$ 0.3 & 3.1 $\pm$ 0.1 & 46.8 $\pm$ 0.3 & 0.062 \\
0.45 & 21.4 & 49.3 $\pm$ 0.3 & 4.1 $\pm$ 0.1 & 46.7 $\pm$ 0.3 & 0.081 \\
0.45 & 17.1 & 49.5 $\pm$ 0.3 & 5.2 $\pm$ 0.1 & 45.3 $\pm$ 0.3 & 0.103 \\
0.45 & 12.8 & 48.6 $\pm$ 0.2 & 7.4 $\pm$ 0.1 & 43.9 $\pm$ 0.2 & 0.144 \\
0.90 & 38.6 & 47.8 $\pm$ 0.3 & 10.1 $\pm$ 0.1 & 42.1 $\pm$ 0.3 & 0.193 \\
0.45 & 8.5 & 47.6 $\pm$ 0.3 & 11.3 $\pm$ 0.2 & 41.1 $\pm$ 0.3 & 0.216 \\
0.90 & 21.4 & 47.3 $\pm$ 0.2 & 11.6 $\pm$ 0.2 & 41.1 $\pm$ 0.2 & 0.220 \\
0.90 & 51.5 & 46.0 $\pm$ 0.3 & 13.9 $\pm$ 0.2 & 40.2 $\pm$ 0.3 & 0.257 \\
1.35 & 30.0 & 45.6 $\pm$ 0.3 & 16.5 $\pm$ 0.1 & 37.9 $\pm$ 0.3 & 0.303 \\
1.35 & 25.7 & 44.9 $\pm$ 0.3 & 16.8 $\pm$ 0.2 & 38.3 $\pm$ 0.3 & 0.305 \\
1.35 & 21.4 & 44.8 $\pm$ 0.3 & 17.6 $\pm$ 0.1 & 37.6 $\pm$ 0.3 & 0.319 \\
0.90 & 8.5 & 45.0 $\pm$ 0.3 & 18.5 $\pm$ 0.2 & 36.5 $\pm$ 0.3 & 0.336 \\
1.35 & 12.8 & 43.1 $\pm$ 0.3 & 20.6 $\pm$ 0.1 & 36.4 $\pm$ 0.3 & 0.361 \\
1.35 & 10.7 & 42.9 $\pm$ 0.2 & 22.3 $\pm$ 0.2 & 34.7 $\pm$ 0.2 & 0.391 \\
1.35 & 8.5 & 42.2 $\pm$ 0.3 & 24.2 $\pm$ 0.2 & 33.6 $\pm$ 0.3 & 0.419 \\
1.35 & 6.3 & 41.9 $\pm$ 0.2 & 26.4 $\pm$ 0.2 & 31.7 $\pm$ 0.3 & 0.454 \\
1.35 & 4.2 & 41.6 $\pm$ 0.2 & 28.6 $\pm$ 0.2 & 29.8 $\pm$ 0.3 & 0.490 

\\
\hline\hline
\end{tabular}
\end{center}
\end{table}
\clearpage{}

\section{XRD}

\begin{figure}[h!]
\begin{center}
\includegraphics[width=\linewidth]{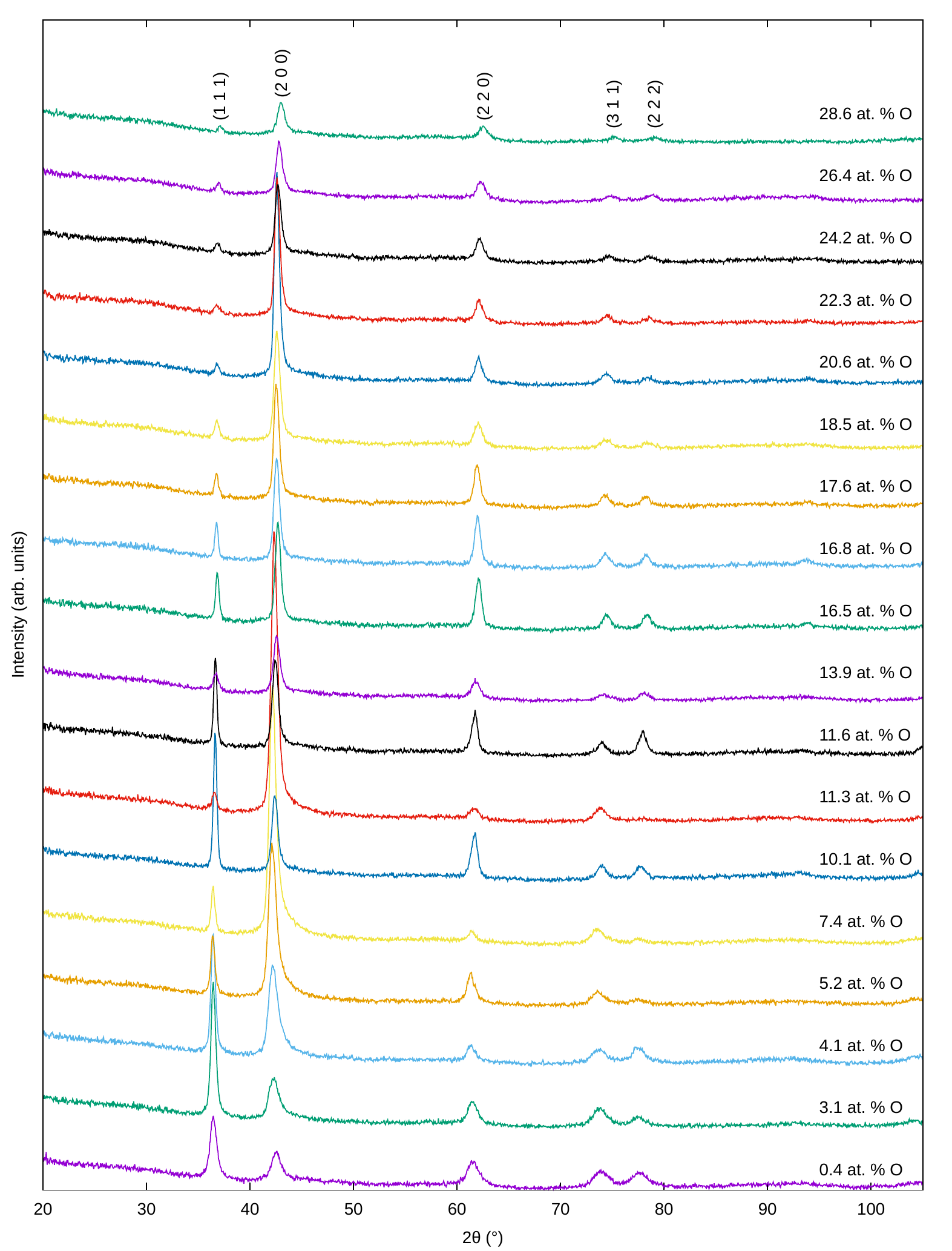}
\caption{\label{fig:XRD-over} XRD diffraction patterns of all samples.}
\end{center}
\end{figure}

\begin{figure}
\begin{center}
\includegraphics[width=\linewidth]{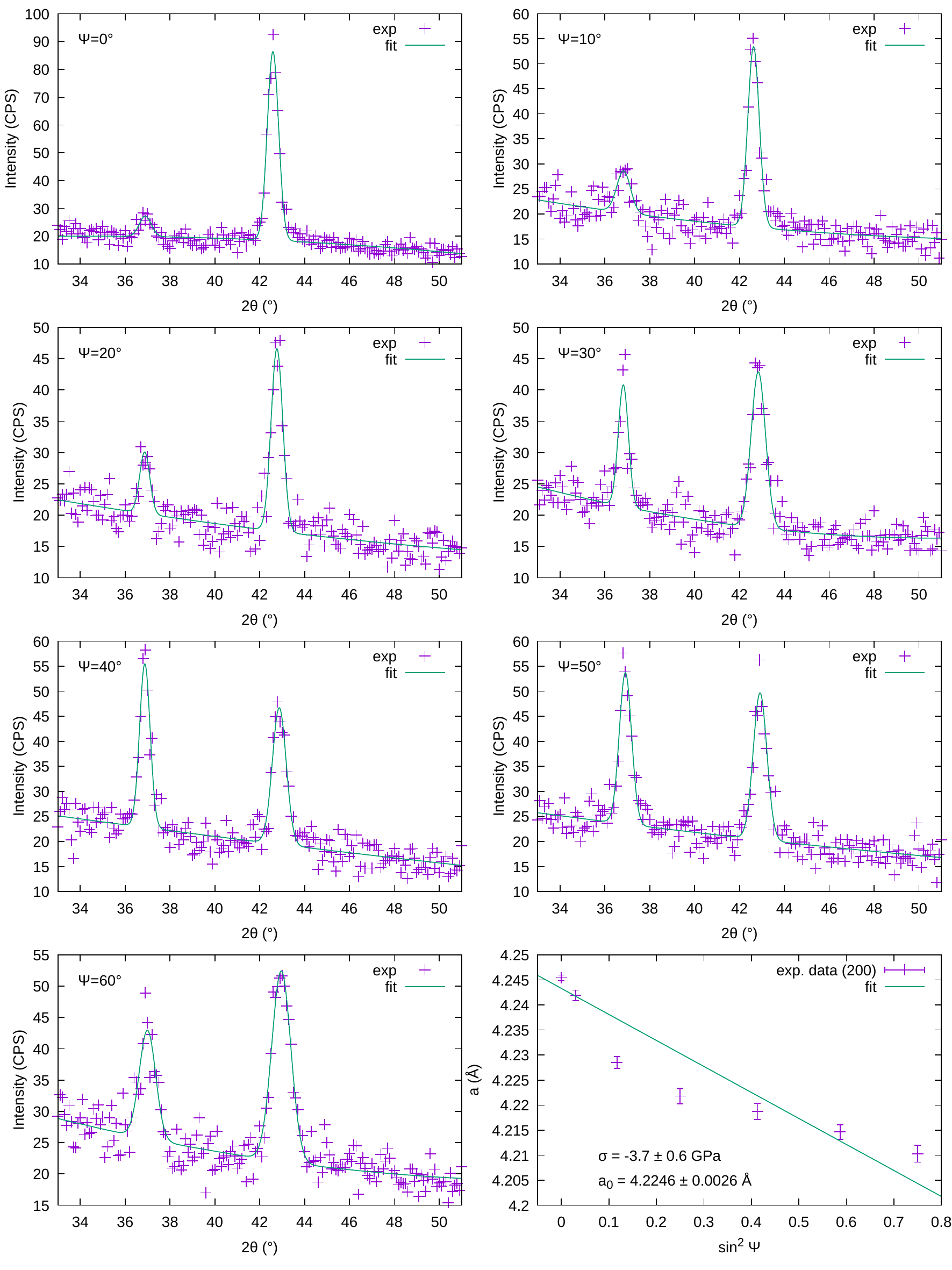}
\caption{\label{sin2_8}Sin$^2\Psi$ fit of the sample with 18.5\,at.\,\%\,O}
\end{center}
\end{figure}

X-ray diffraction patterns were measured at Bruker D8 Discovery general area detector diffraction system (GADDS) using Cu(K$_\alpha$) radiation with a wavelength $\lambda =$ 1.54\,Å at a fixed incidence angle of 15$^\circ$.
The XRD spectra were constructed from three scans with the area detector centered at 30, 60 and 90$^\circ$.
Independent of the oxygen concentration, all samples exhibit a single cubic phase in the XRD, see Figure~\ref{fig:XRD-over}.
The equilibrium lattice parameters, were determined using the sin$^2\Psi$ method~\cite{Cullity1978}
\begin{equation}
a(\Psi) = \frac{1+\nu}{E} \sigma a_0 \sin^2\Psi + a_0 \left(1- \frac{2\nu\sigma}{E}\right) \, ,
\end{equation}
where $a_0$, $E$, $\nu$ and $\sigma$ are equilibrium lattice parameter, elastic modulus, Poisson's ratio and film stress, respectively.
For this task, the samples were measured using GADDS in a Bragg-Brentano geometry centered on the (200) diffraction and seven different $\Psi$ angles were scanned from 0 to 60$^\circ$.
As sin$^2\Psi$ method requires elastic modulus and Poisson's ratio as an input, experimental results reported previously~\cite{Do2014} were interpolated by the means of polynomial fit
to get estimates for the specific oxygen concentrations corresponding to our samples.
The resulting fitted dependence of the elastic modulus on the oxygen content is a follows
\begin{equation}
E([\mathrm{O}]) = (1465.22 [\mathrm{O}]^2 -823.81 [\mathrm{O}] + 473.54 ) \, \mathrm{GPa} .
\end{equation}
We however note that the influence of the used elastic modulus on the determined equilibrium lattice parameter is small as it mostly influences the determined stress,
which is however not discussed here.
Example fit of the peak positions at the different $\Psi$ angles and the final sin$^2\Psi$ fit is shown in Figure~\ref{sin2_8}.

The comparison of absolute values of experimental equilibrium lattice parameter at 300\,K with the 0\,K DFT results is challenging.
Besides the difference caused by thermal expansion,
PBE is known to usually overestimate the lattice parameter~\cite{Zhang2018}.
While it is possible to go beyond the PBE, or to estimate the DFT lattice parameters at higher temperatures,
this is not within the scope of this work. Therefore, we only employ a simple correction.
The reported lattice parameter for bulk TiN of 4.241\,\r{A}~\cite{Lengauer1992} is extrapolated to 0\,K value of 4.238\,\r{A} using the calculated thermal expansion coefficient of TiN by Bartosik et al.~\cite{Bartosik2017}.
This is compared to our PBE lattice parameter of 4.249\,\r{A} for TiN and the difference is 0.011\,\r{A}.
We further assume for simplicity that the PBE overestimation and the thermal expansion is constant in the whole composition range and
this constant correction is applied to all of the calculated DFT lattice parameters.

\begin{figure}[h!]
\begin{center}
    \includegraphics[width=0.8\linewidth]{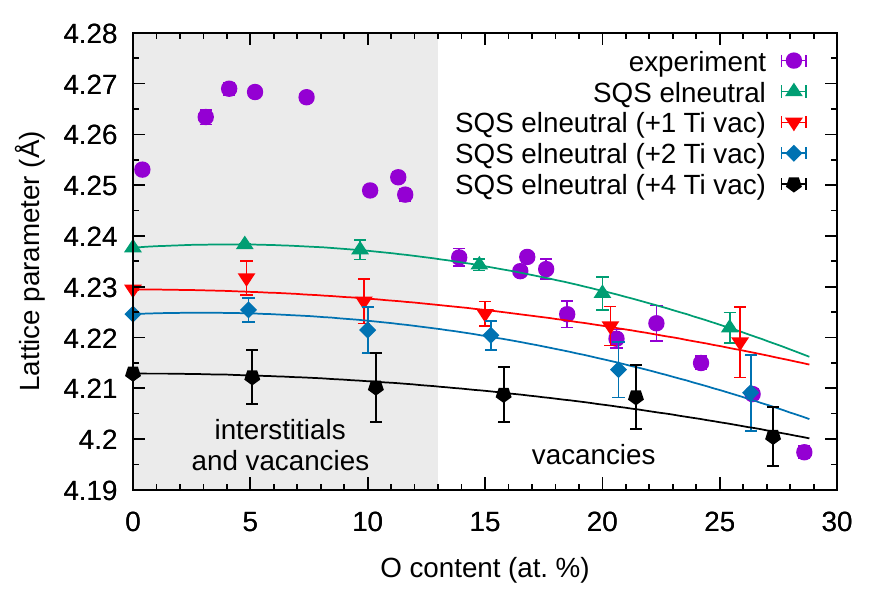}
    \caption{\label{fig:latpar}Equilibrium lattice parameter of the c-\TiON{} thin films and of the DFT models.}
    \end{center}
\end{figure}

The experimental stress-free lattice parameters as determined by XRD are shown in Figure~\ref{fig:latpar}, together with the lattice parameter as obtained from DFT models.
The DFT models show a steady decrease of the lattice parameters in the whole compositional range with the models containing more vacancies corresponding to smaller lattice parameter.
However, the experimental lattice parameter at 0.4\,at.\,\%\,O (4.253\,\r{A}) is larger than the reported value for bulk TiN (4.241\,\r{A}~\cite{Lengauer1992}) and further increases up to 5--7\,at.\,\% O, before continuously decreasing.
This initial overestimation of the lattice parameter can be explained by the presence of non-metal interstitials~\cite{Baben2013}.
It was previously shown that an O interstitial in TiN has a formation energy of just 0.69\,eV,
while the formation energy of various types of N interstitials was reported in the 3.4-5.4\,eV range~\cite{Baben2013, Tsetseris2007}.
So while the high value of lattice parameter at 0.4\,at.\,\%\,O is likely caused by N interstitials, 
the further increase in lattice parameter up to 5--7\,at.\,\% O is attributed to the energetically favored 
interstitial incorporation of O as compared to N~\cite{Baben2013}, facilitated by ion bombardment~\cite{KarimiAghda2021}.
Starting at 10--15\,at.\,\% O content, a continuous decrease in lattice parameter is visible, here attributed to Ti vacancies.
Therefore, it is inferred, that as the O content increases, the role of the non-metal defects decrease and the Ti vacancies are becoming the dominant defect.

\begin{figure}[h!]
\begin{center}
\includegraphics[width=0.8\linewidth]{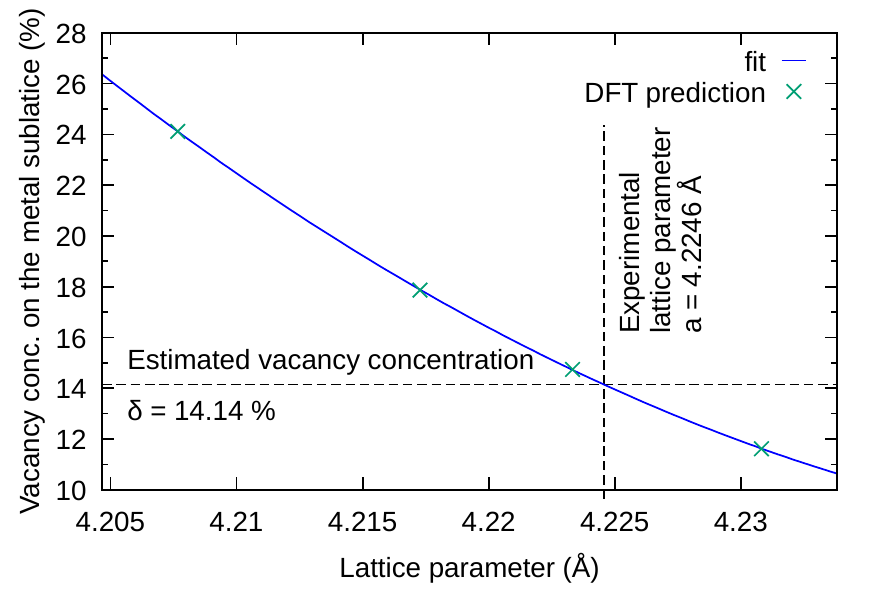}
\caption{\label{fig:XRD-fit} Vacancy concentration estimate for the sample with 18.5\,at.\,\% O based on the DFT and experimental lattice parameters.}
\end{center}
\end{figure}

Employing the DFT lattice parameter predictions and under the assumption of Ti vacancies being the only defect present in the films,
Ti vacancy concentration estimates were made from the experimental lattice parameter.
The calculated mean lattice parameters for the SQS cells
were interpolated with a second order polynomial to get the expected lattice parameter for an electroneutral vacancy scenario as well as for the scenarios with more vacancies.
The fits are shown as full lines in the Figure~\ref{fig:latpar}.
At a later point, for a given specific O content of every sample, the vacancy concentration on the metal sublattice corresponding to the
interpolated DFT models was plotted as function of the lattice parameter and fitted with a second order polynomial,
see Figure~\ref{fig:XRD-fit} for one selected composition as an example.
The experimental vacancy concentration estimate is determined as the value of this fitted function at the experimental lattice parameter of the specific sample.
i.e, by interpolating between the calculated lattice parameters for different vacancy concentrations at the constant O content
to predict the vacancy concentration corresponding to the specific experimental lattice parameter.

\clearpage{}

\section{Local chemical composition and microstructure}

A site-specific lift-out for microstructural characterization was carried out by focused ion beam techniques in a Helios Nanolab 660 (FEI Company) dual-beam microscope.
Lamellae were extracted in the growth direction of the thin films and Ga ions were used at an acceleration voltage of 30\,kV. Final lamella thicknesses were $<$ 100\,nm.
The film microstructure was subsequently investigated in scanning transmission electron microscopy (STEM) mode at 30\,kV using a STEM III detector.
Three-dimensional distribution of constitutional elements was investigated by atom probe tomography (APT) using a CAMECA LEAP 4000X HR.
Field evaporation was assisted by laser pulsing at 200\,kHz repetition frequency,
while the base temperature was 60\,K and the detection rate was set to 0.5\,\%.
The laser pulse energy of 10\,pJ was employed in order to increase the electric field strength
and enhance the ionization probability of neutral species~\cite{Hans2020}.
Data reconstruction was performed with IVAS (version 3.8.0) using the shank angle reconstruction protocol.

\begin{figure*}
    \includegraphics[width=\linewidth]{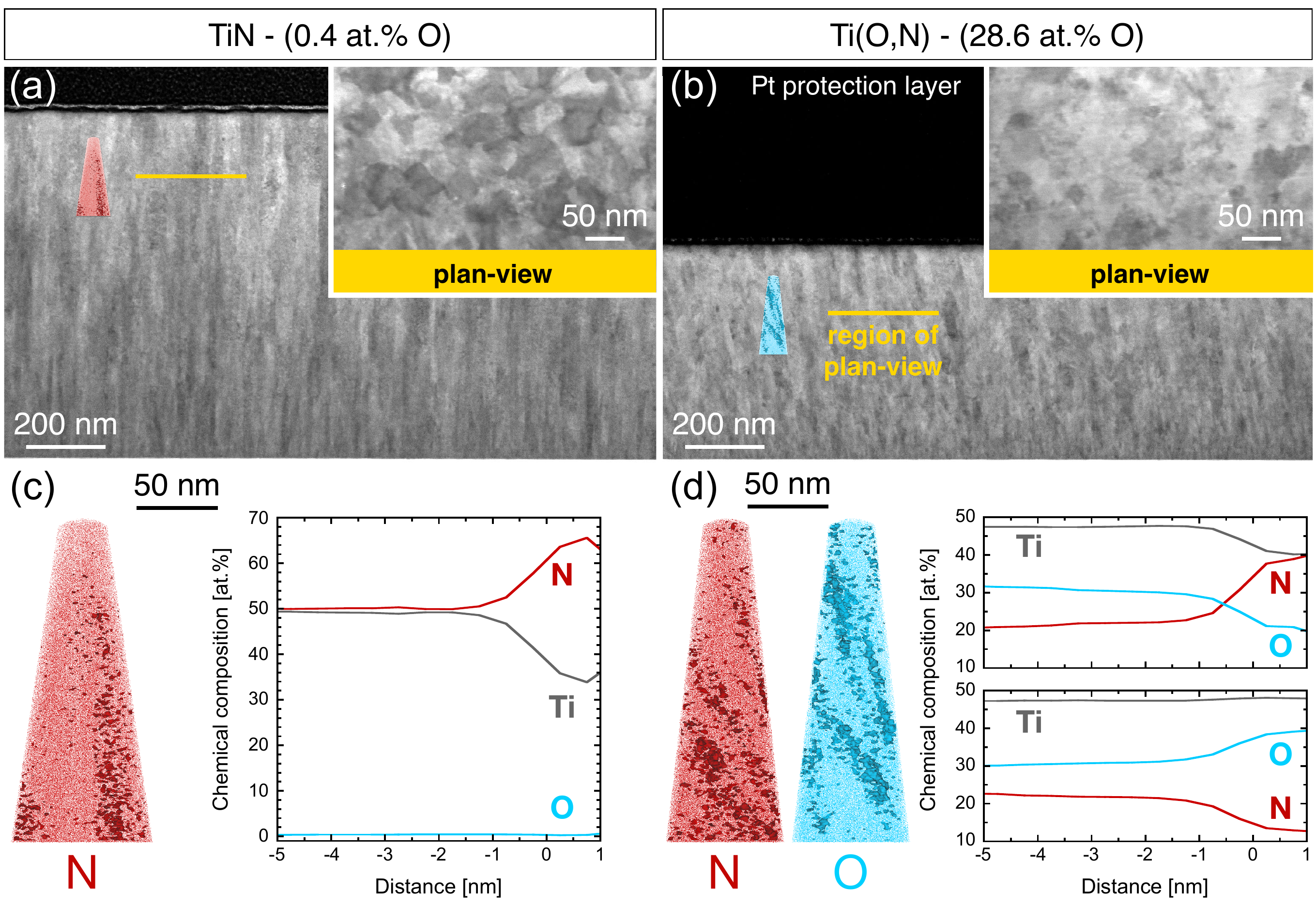}
    \caption{\label{fig:APT}STEM images of films with the (a) lowest (0.4\,at.\,\% O) (b) highest O content (28.6\,at.\,\% O) and APT images of the (c) N rich regions in sample with 0.4\,at.\,\% O and (d) N and O rich regions in sample with 28.6\,at.\,\% O.}
\end{figure*}

Bright field images of the thin films with 0.4 and 28.6\,at.\,\% are shown in Figure~\ref{fig:APT}~(a) and (b), respectively.
The thickness reduction from approximately 880 (TiN, 0.4 at.\,\% O) to 540 nm (TiON, 28.6\,at.\,\% O) can be understood based on the oxygen-induced decrease of the sputtering yield~\cite{Severin2006}.
 While TiN exhibits a dense microstructure with columnar grains~\cite{Petrov2003},
 O addition causes the formation of a fine-grained nanocrystalline microstructure,
 consistent with XRD results.
The local chemical composition is investigated by APT, Figure~\ref{fig:APT}~(c) and (d).
Atomic positions of N are shown for the thin film with 0.4\,at.\,\% O in Figure~\ref{fig:APT}~(c) and N-rich regions are highlighted by isoconcentration surfaces with $\ge$ 54\,at.\,\%\,N.
Local clustering of N can observed with enrichments of up to 65\,at.\,\% while Ti is depleted to approximately 34\,at.\,\%.
For the thin film with 28.6\,at.\,\% O,
atomic positions of TiN and TiO molecular ions are displayed in Figure~\ref{fig:APT}~(d).
In TiN-rich regions a significant O reduction from 30 to 18\,at.\,\% is observed,
while Ti is reduced from 48 to 40\,at.\,\%.
In case of TiO-rich regions the Ti content is constant with 47\,at.\,\%, while O increases from 28 to 36\,at.\,\%
and N is reduced from 26 to 17 at.\,\%.
It is most likely that the chemical composition modulations pertain to the grain boundaries,
i.e., N  and O-rich regions are at grain boundaries.
This could be explained by excess of O and N which cannot
be incorporated into the film and segregate at the grain boundaries~\cite{Petrov2003}.

\clearpage{}

\section{XPS}

\begin{figure}[h!]
\begin{center}
\includegraphics[width=0.9\linewidth]{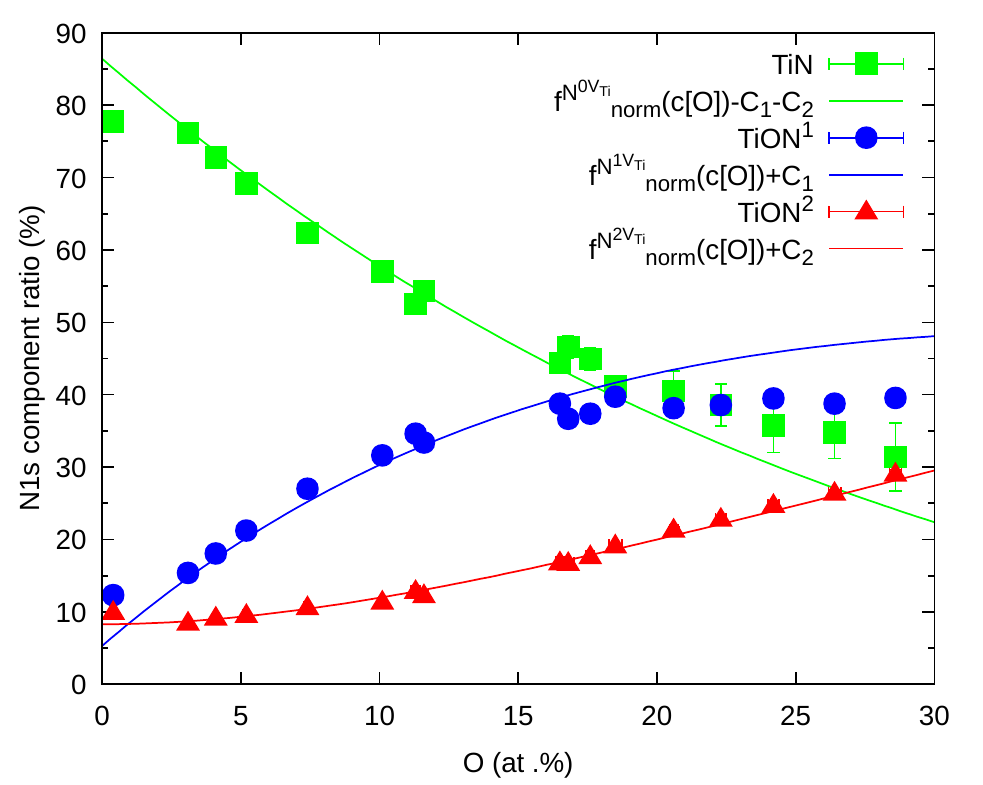}
\caption{\label{fig:N1s-const-fit} Fit of the $C_1$ and $C_2$ constants corresponding to the additional oxidation after atmospheric exposure.}
\end{center}
\end{figure}

\begin{figure}[h!]
\begin{center}
\includegraphics[width=\linewidth]{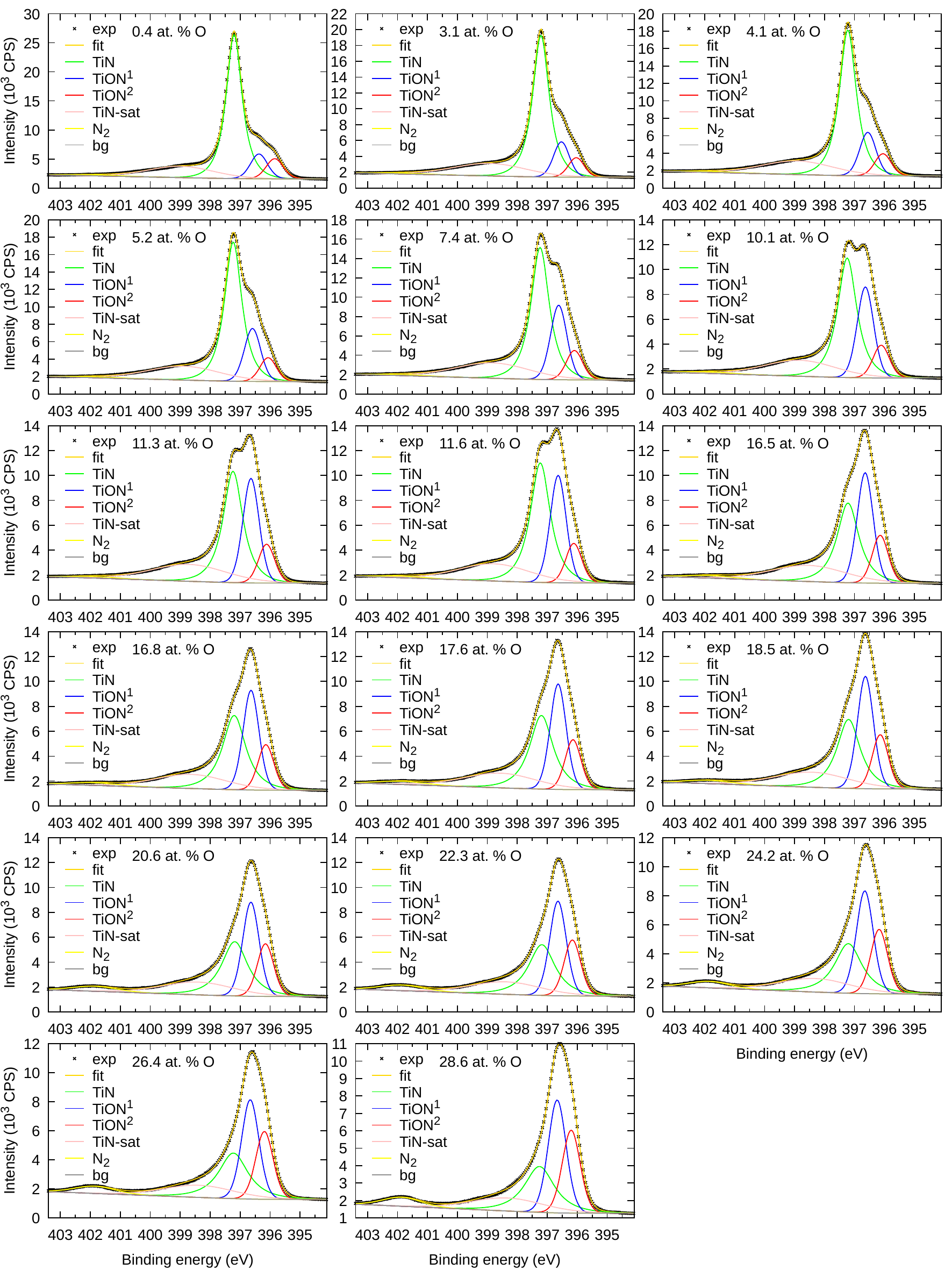}
\caption{\label{fig:N1s-over} XPS N\,1s fits of all samples.}
\end{center}
\end{figure}

Figure~\ref{fig:N1s-const-fit} shows the fit of the N\,1s components to obtain the $C_1$ and $C_2$ constants and Figure~\ref{fig:N1s-over} shows the XPS N\,1s region fits for all compositions.

\clearpage{}

\section{Discussion of the Ti vacancy estimates by the different methods}

From Figure~6 in the manuscript it is evident that the vacancy expectations based on the electroneutral model are consistent
with the here proposed model utilizing binding energy data obtained from the N \,1s core spectra region via XPS over the whole composition range.
In addition, Figure~6 also depicts the oxygen content-dependent predictions
of Ti vacancy concentration based on XRD, which were obtained through the approach described in Section~S4.

The prediction replicates the trend of the vacancy expectations based on the electroneutral model
and hence the XPS based model communicated here.
However, composition dependent deviations are also observed, where the negative vacancy concentrations
at low O concentrations (Figure 6) stand out.
These XRD data-based negative vacancy concentrations in the O concentration range of up to $\sim$15\,at.\,\%
 may be rationalized by considering the effect of O and N interstitials.
For the XRD vacancy quantification method, where the vacancy concentration is estimated solely based on the lattice parameter,
the non-metal interstitials are causing an increase in lattice parameter.
Therefore, the method can result in nonphysical negative Ti vacancy concentration predictions
when used for samples with interstitial.
The assumption of Ti vacancies being the only defect present in the film is hence being questioned for films with low O content.

The effect of possible interstitials on the XPS N\,1s BEs and therefore on the vacancy quantification was not studied in this work.
However, contrary to the XRD method,
the XPS-based vacancy estimates at the low O content are in good agreement with the electroneutral model.
Therefore, it is reasonable to assume that the binding energies shifts associated with the non-metal interstitials are too small to be detected
and the interstitials are therefore undetectable in the XPS.
Another possibility is that the effect of interstitials was subtracted together with the correction of the surface oxidation influence. 
The exact effects of the different defect types besides just vacancies on the binding energies will be investigated in a follow up work.

At O contents $>$ 18\,at.\,\%, the XRD-based vacancy estimation results in an overestimation compared to both the XPS-based method and the electroneutral model prediction.
The high XRD vacancy estimates,
most visible for the sample with the highest O concentration, could originate from the assumptions utilized therein.
For example, the assumption of constant thermal expansion in the whole O range could at least partially explain the apparent higher Ti vacancy concentration at higher O contents.
Another shortcoming of this method is that the calculated lattice parameter is quite sensitive to
the random arrangements between the SQS-populated metal and non-metal sublattices at high O concentrations,
as indicated by the error bars showing the standard deviation of the values in Figure~\ref{fig:latpar} for the three SQS cells.
This suggests that the relation between the non-metal and metal sublattice ordering is in fact non-negligible for the estimation of the lattice parameter.
Examining the DFT structures show that for O contents $>$ 25\,at.\,\%, the cubic-like structure even
starts to collapse locally if the local O and Ti vacancy concentrations are too high.
This probably does not occur in the films as there is an energy penalty for vacancy clusters~\cite{Tsetseris2007}.
Therefore, the SQS-based structures might not provide good estimates for the lattice parameter trends at high O contents.
The DFT trends are however used as a basis for the XRD vacancy estimates,
so any DFT lattice parameter uncertainty translates directly to an uncertainty
of the vacancy estimates based on the experimental lattice parameters as determined by XRD.
Importantly, the non-metal and metal sublattice ordering plays a minor role for the BE quantification.
This is demonstrated in Figure~\ref{fig:spectra} where the differences between the N\,1s broadened BEs for different structures at the same composition are minimal.
This is consistent with the notion that the N\,1s BEs are influenced only by the local environment and not the global cell arrangement.

Another possible explanation of the discrepancy at high O contents can be found in the APT data, which indicate some inhomogeneity in the N and O distribution.
The XPS quantification method is based on the analysis of N 1s signal,
therefore, the measured XPS spectra will give more weight to the N-rich regions.
The inhomogeneity could be affecting the measured XRD lattice parameter (and hence the XRD-based vacancy concentration estimates) as well,
if the N- and O-rich regions have different crystallinity.
Hence, it is evident that the here proposed XPS based method to quantify the Ti vacancy concentrations is superior to the XRD-based method.

\bibliographystyle{elsarticle-num}
\bibliography{bib-db}